\documentclass[a4paper,11pt]{article}
\pdfoutput=1 % if your are submitting a pdflatex (i.e. if you have
\usepackage{jinstpub} % for details on the use of the package, please
\usepackage{graphicx}
\usepackage{subfigure}

\title{\boldmath Readout electronics and data acquisition system of PandaX-4T experiment}

\author[a]{Jijun Yang,}
\author[a]{Xun Chen,}
\author[a]{Changda He,}
\author[a]{Di Huang,}
\author[b]{Yanlin Huang,}
\author[a,c]{Jianglai Liu,}
\author[d,1]{Xiangxiang Ren,\note{Corresponding author}}
\author[d]{Anqing Wang,}
\author[d]{Meng Wang,}
\author[a]{Binbin Yan,}
\author[a]{Kai Yin,}
\author[a]{Jinqun Yang,}
\author[a,2]{Yong Yang,\note{Corresponding author}}
\author[b]{Qibin Zheng} 

\affiliation[a]{School of Physics and Astronomy, Shanghai Jiao Tong University, MOE Key Laboratory for Particle \\
  Astrophysics and Cosmology, Shanghai Key Laboratory for Particle Physics and Cosmology, 200 Dongchuan Road, Shanghai 200240, China}
\affiliation[b]{Institute of Biomedical Engineering, University of Shanghai for Science and Technology, 516 Jungong Road, Shanghai 200093, China}
\affiliation[c]{Tsung-Dao Lee Institute, 520 Shengrong Road, Shanghai 200240, China}
\affiliation[d]{Key Laboratory of Particle Physics and Particle Irradiation (MOE), Institute of Frontier and Interdisciplinary Science, Shandong University, 72 Binhai Road, Qingdao, Shandong 266237, China}

\emailAdd{renxx@sdu.edu.cn}
\emailAdd{yong.yang@sjtu.edu.cn}

\abstract{PandaX-4T is a dark matter direct detection experiment
  located in China jinping underground laboratory. The central
  apparatus is a dual-phase xenon detector containing 4 ton liquid
  xenon in the sensitive volume, with about 500 photomultipliers
  instrumented in the top and the bottom of the detector. In this
  paper we present a completely new system of readout electronics and
  data acquisition in the PandaX-4T experiment. Compared to the one
  used in the previous PandaX dark matter experiments, the new system
  features triggerless readout and higher bandwidth. With triggerless
  readout, dark matter searches are not affected by the efficiency
  loss of external triggers. The system records single photelectron
  signals of the dominant PMTs with an average efficiency of 96\%, and
  achieves the bandwidth of more than 450 MB/s. The system has been
  used to successfully acquire data during the commissioning runs of
  PandaX-4T.}


\begin{document}
\maketitle
\flushbottom

\section{Introduction}
The existence of dark matter has been established by many
cosmological and astrophysical observations. Detection of DM
particles is the central goal of many experiments. The PandaX project
consists of a series of xenon-based experiments to search for DM
particles and to study the properties of neutrinos, utilizing the time
projection chamber (TPC) technique. PandaX-4T is a 4 ton scale
dual-phase TPC, located in the B2 hall of the second phase of China
jinping underground laboratory (CJPL-II). Particle interaction in the
liquid xenon produces a prompt scintillation signal (S1) and
ionization electrons which are drifted up to the gaseous xenon region. These
electrons subsequently produce a delayed scintillation signal
(S2). The top and the bottom of the TPC are instrumented with 169 and
199 3-inch Hamamatsu R11410-23 photomultipliers (PMTs), Two circles of
105 1-inch Hamamatsu R8520-406 PMTs are placed at the top and the
bottom of the veto compartment to reduce peripheral background.

The main goal of the electronics and data acquisition system is to
digitize the electrical signals from PMTs and record data that might
correspond to physical interactions in the liquid xenon. With the upgrade of the
detectors in PandaX, the readout electronics and DAQ system have also evolved. In previous
experiments PandaX-I and PandaX-II~\cite{Ren:2016ium, Wu:2017cjl}, the
data acquisition of digitizers (CAEN V1724, 100 MS/s sampling rate)
relies on global external trigger signals. The trigger system is
designed to trigger on relatively large S2 signals.  Upon a trigger
request, the data of each channel during a fixed time window are
read out with baseline data suppressed in the digitizers. The length of
the window is usually set to be at least twice of the maximum electron
drift time in the liquid phase, in order to make sure both S1 and S2
signals are recorded.

This global-trigger based data acquisition scheme is straightforward. A trigger defines a possible physical event. However, the
sensitivities of DM searches are inevitably affected by the trigger
efficiency loss.  Many DM-xenon interactions, including weakly
interacting massive particles (WIMP)-nucleon interactions, DM-nucleon
interactions via a light-mediator, light DM-electron scatterings, and
so on, produce S2 signals with a roughly exponentially falling
distribution. Especially, the expected signals from light-mediator DM
models~\cite{Ren:2018gyx,Yang:2021adi} and light DM
models~\cite{Cheng:2021fqb} are more peaked towards the trigger
threshold than that of the WIMP model. For example, in the light
DM-electron analysis~\cite{Cheng:2021fqb}, the S2 lower cut is set to
be the same as the trigger threshold. In this case, trigger
inefficiency becomes one of the dominant causes for signal detection
efficiency loss.

In PandaX-4T, the readout electronics system is designed with new
digitizers (CAEN V1725, 250 MS/s). This digitizer is capable of
acquiring data without relying on the global triggers. Each digitizer
channel can be self-triggered independently. In this case, each
self-trigger corresponds to the PMT likely being hit by scintillation
lights. The data of each channel are then read out together with the
self-trigger time information. Besides the new digitizer system, the
DAQ system is also redesigned with improved bandwidth. In PandaX-I and
PandaX-II, several digitizers are daisy chained and their data are
read out via an optical link. Only one DAQ readout server is used.  In
PandaX-4T, data of each digitizer are read out through a separated
optical link. Four DAQ readout servers are used.  Data from each
server are sent to another server for further processing and finally
saved into the disk.

Table~\ref{pandax_daq} summarizes the evolvement of the electronics and DAQ system 
from PandaX-I to PandaX-4T. In this paper we present the PandaX-4T electronics and DAQ system and its
performance in the commissioning runs. We only discuss performance related to the dominant 3-inch PMTs.

\begin{table}
\begin{center}
\begin{tabular}{ |p{2cm}||p{2cm}|p{3cm}|p{2.5cm}|p{2.5cm}|  }
 \hline
% \multicolumn{5}{|c|}\\ \hline
 Experiment & Number of PMTs & Digitizers & Trigger and DAQ & Data transmission\\ \hline
 PandaX-I & 183 & V1724 (100MS/s) & Global-trigger based & Daisy chained\\ \hline
 PandaX-II & 158 & V1724 & Global-trigger based, upgraded & Daisy chained\\ \hline
 PandaX-4T & 473 & V1725 (250MS/s) & Triggerless & Parallel\\ \hline
\end{tabular}
 \end{center}
  \caption{Evolvement of the electronics and DAQ system in PandaX project from
    PandaX-I to PandaX-4T. The same system is used in PandaX-I and
    PandaX-II. The trigger system was upgraded in 2017, reducing the
    threshold from 4 electrons to 2.5 electrons~\cite{Wu:2017cjl}. PandaX-4T
    features a triggerless and higher-bandwidth DAQ.}
  \label{pandax_daq}
\end{table}

\section{Readout Electronics and DAQ of PandaX-4T}
\label{hardware}

\begin{figure*}[!htbp]
  \centering \includegraphics[width=0.8\linewidth]{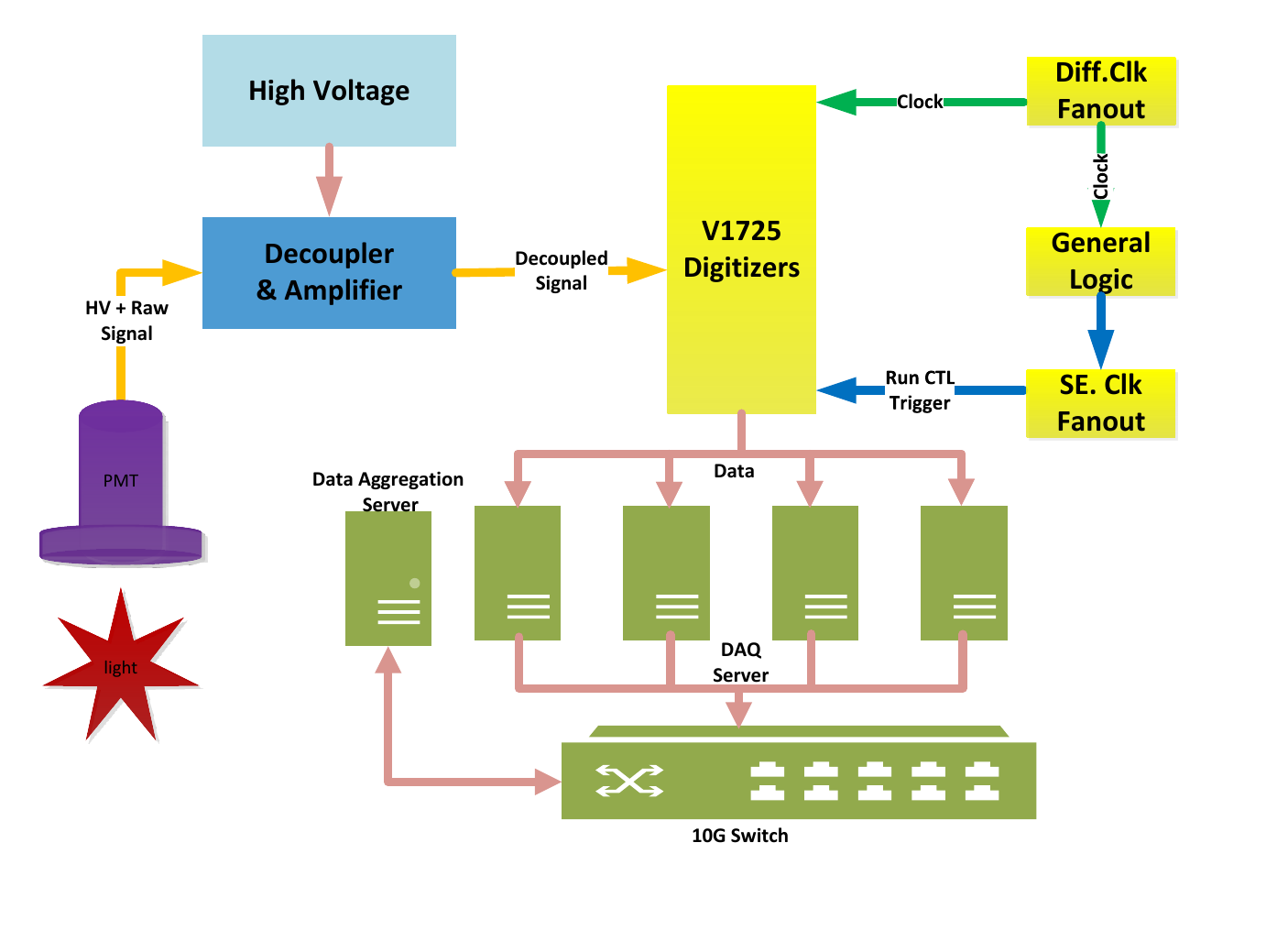}
  \caption{The schematic drawing of PandaX-4T readout electronics and
    DAQ hardware system. See text for details.}
  \label{daq_hd}
\end{figure*}

\begin{figure*}[!htbp]
  \centering \includegraphics[width=0.8\linewidth]{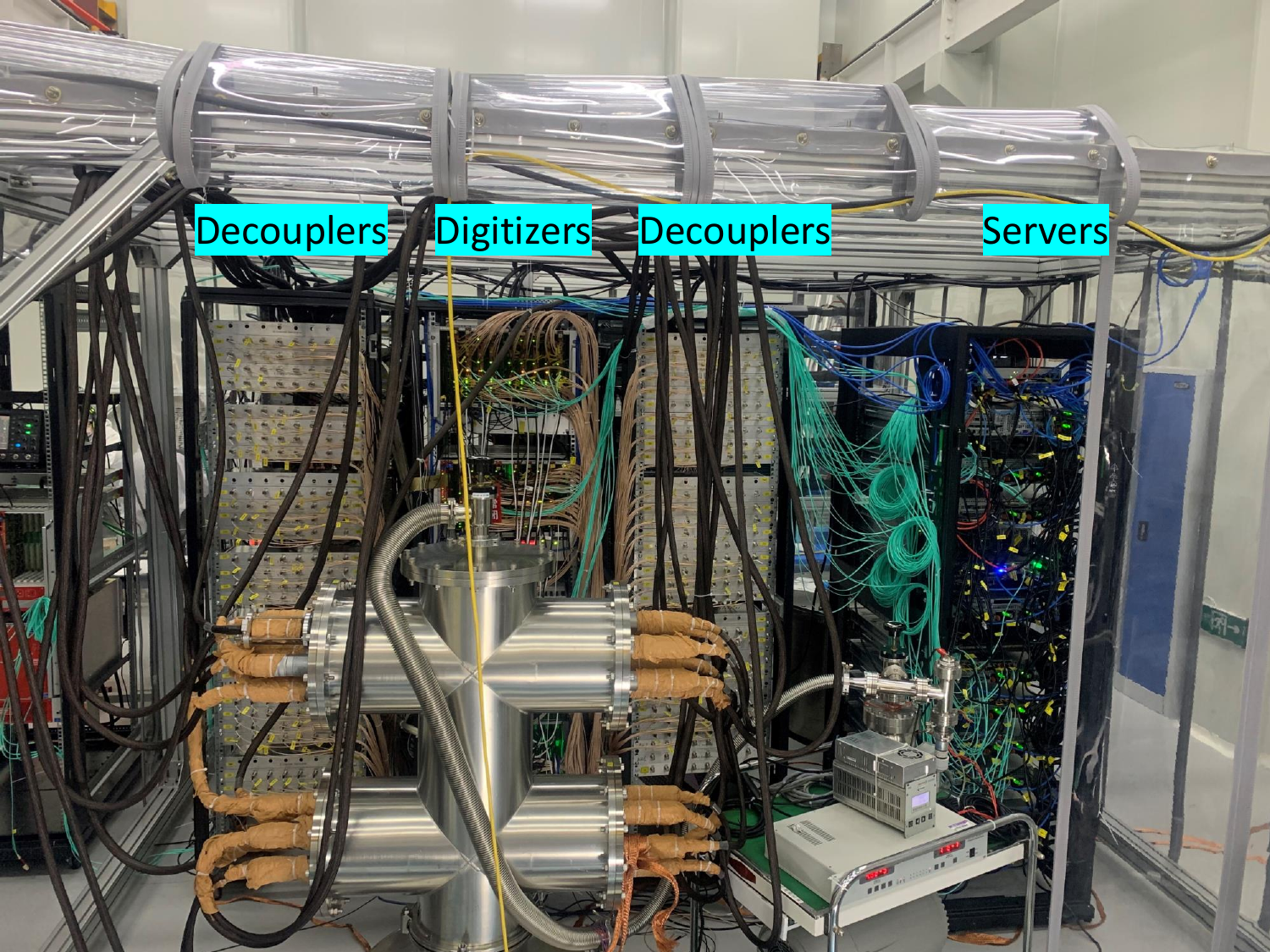}
  \caption{The PandaX-4T readout electronics and DAQ system in
    situ. This system includes two VME crates for the new decouplers,
    one VME crate for the digitizer array, and one VME crate for servers.}
  \label{daqphoto}
\end{figure*}

Figure.~\ref{daq_hd} illustrates the overall design of the PandaX-4T
readout electronics and DAQ system. The system in situ is shown in
Figure~\ref{daqphoto}.

\begin{figure*}[!htbp]
  \centering
  \includegraphics[width=0.8\linewidth]{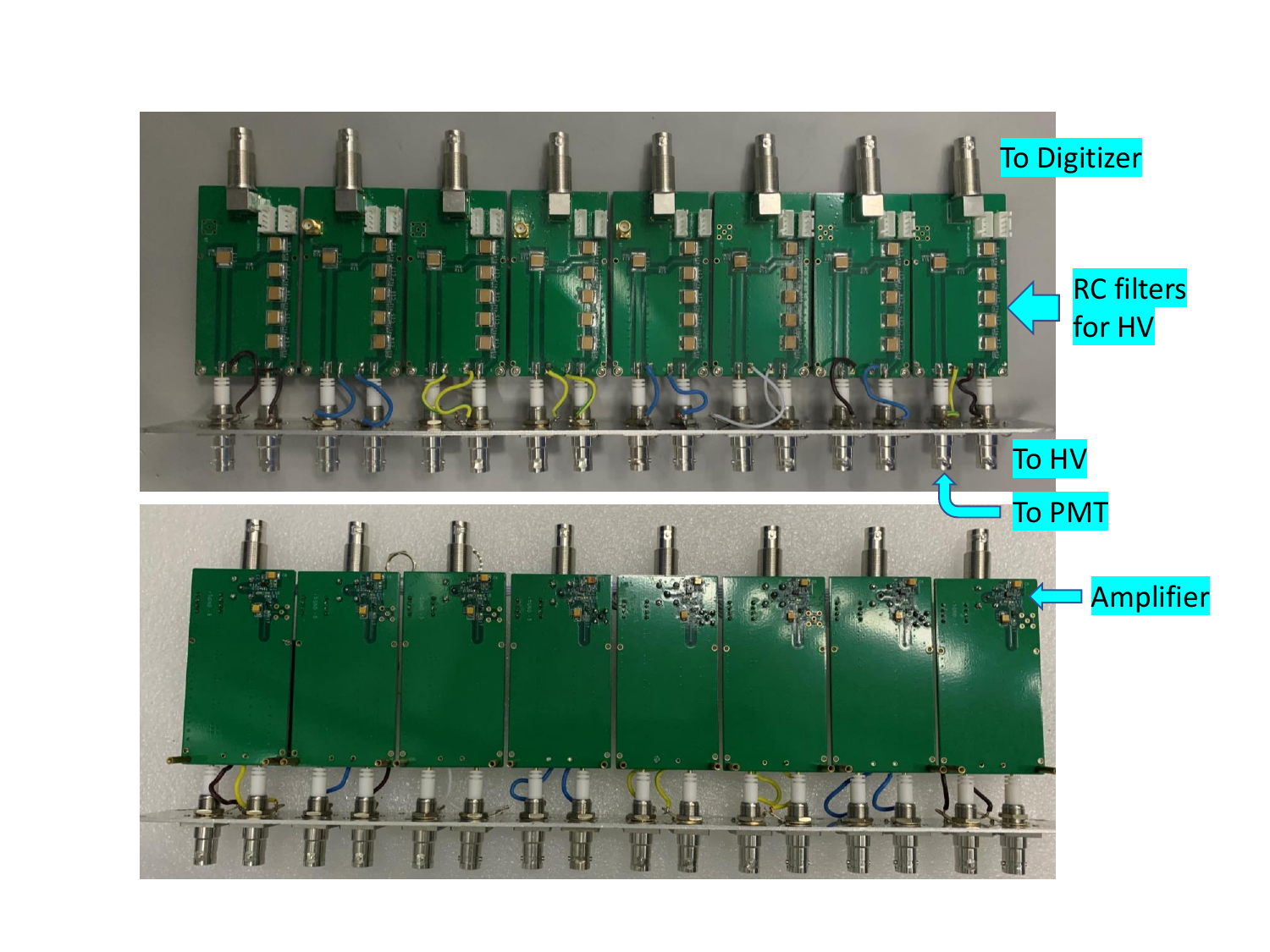}
  \caption{The new decoupler module used in PandaX-4T. It integrates the functionalities
    of PMT HV-signal decoupling and amplification. }
  \label{dcphoto}
\end{figure*}

The whole system consists of three main parts. The first part includes
custom designed PMT signal decoupling and amplification modules. The
anode signal from each PMT is transmitted to outside on the same
coaxial cable which provides the PMT High Voltage (HV) bias. For future development,
14 PMTs are mounted on the new base boards~\cite{Zheng:2020kfp}, which have a new configuration of
decoupling capacitors near the last several dynodes to extend the dynamic range
and provide two readout signals at the anode and the eighth dynode.  These PMTs are located in the
central top and bottom PMT arrays. Their dynode signals are also transmitted to outside using coaxial cables.
Each PMT signal from the anode or the dynode is
separated from the HV via a capacitive decoupling circuit, which is
the same as the one used in previous PandaX experiments.  The
decoupled signal is then amplified through a high-speed operational
amplifier (ADI AD8009). This amplifier has very high slew rate (5.5 V/ns) and large
bandwidth (700 MHz, G=2 at -3dB for small signals)~\cite{adi}. Thus
the impact on the PMT signal shape is expected to be small. The
decoupling and the amplification circuits are integrated on the same
PCB module (called new decouplers below), shown in
Figure~\ref{dcphoto}. The amplification is configured to be 1.5 and 5 for 3-inch PMTs
and 1-inch PMTs, respectively. This new design greatly reduces the
complexity of the whole electronics system. In previous PandaX
experiments, the decouplers and the amplifiers (Phillips 779 NIM
modules) are located in different places and they must be connected
through coaxial cables.

\begin{figure}[!htbp]
  \centering \includegraphics[width=0.48\textwidth]{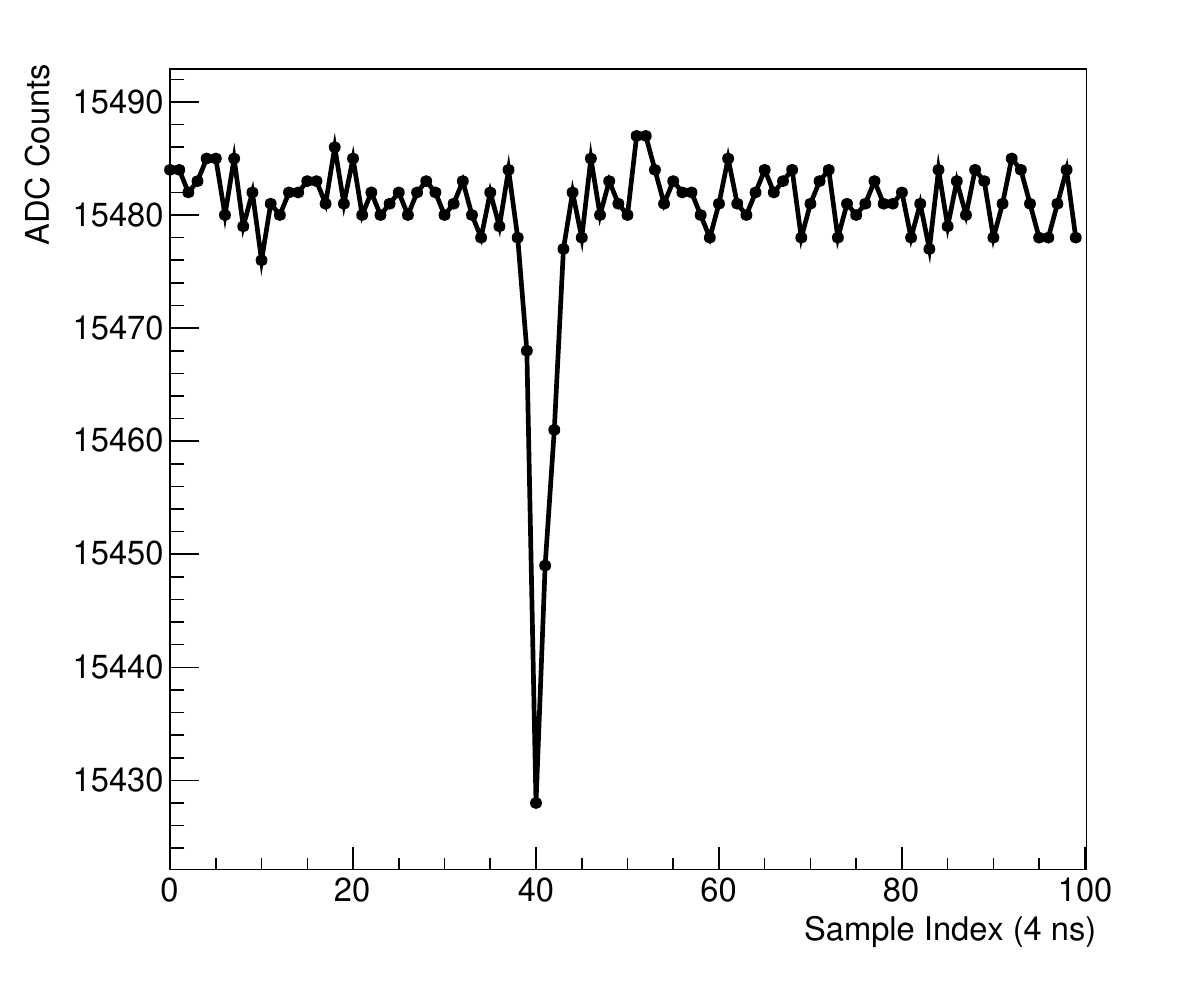}
  \includegraphics[width=0.48\textwidth]{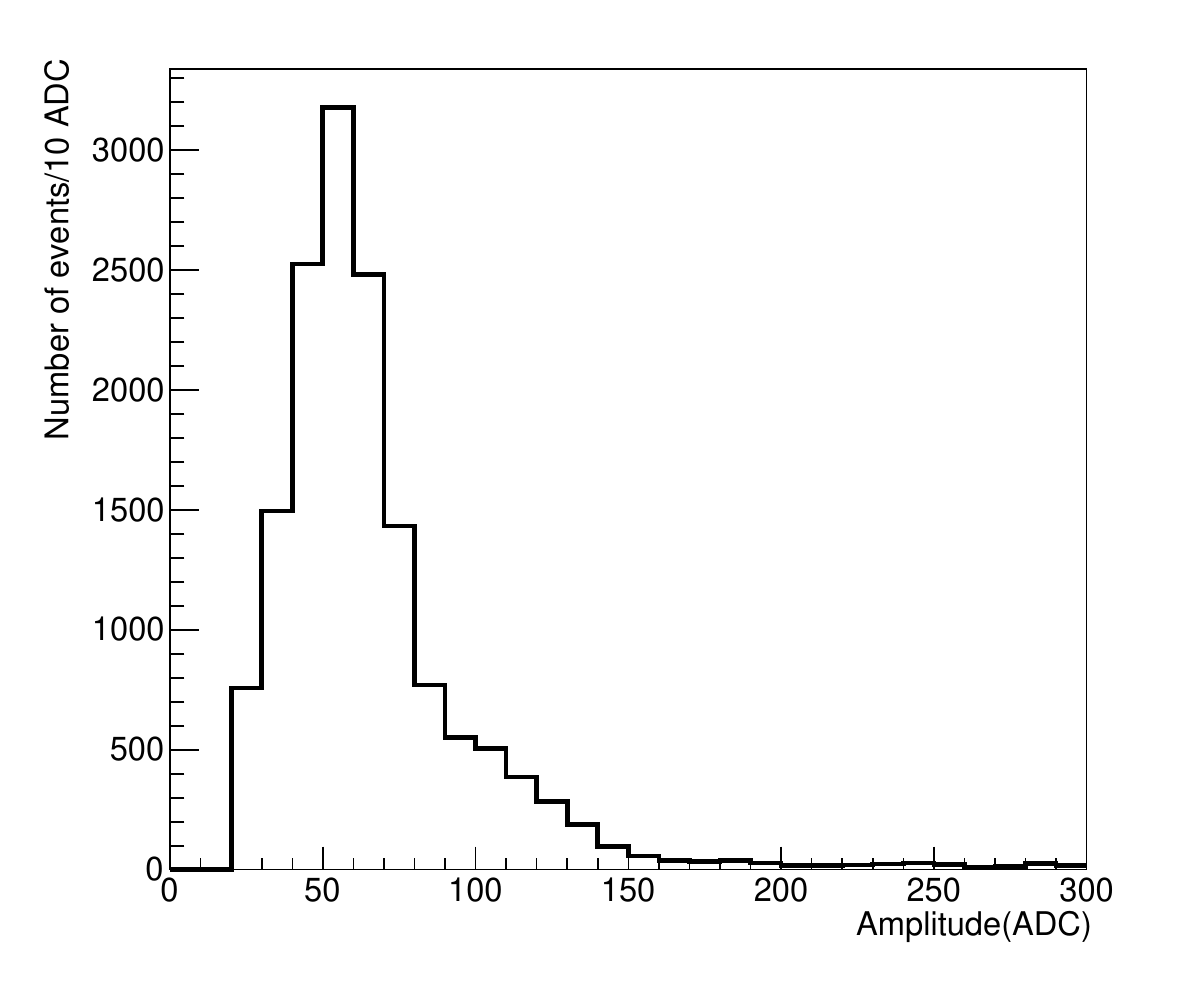}
  \caption{Left, a digitized waveform of a typical SPE signal from a
    3-inch PMT in PandaX-4T.  Right, recorded waveform amplitude distribution of
    one 3-inch PMT channel with a self-trigger threshold of 20 ADC
    counts.}
  \label{spe_wf}
\end{figure}

The second part includes 32 CAEN V1725 digitizers. V1725 is a
16-channel VME module. Each input PMT signal is digitized by a flash ADC
channel with 14-bit resolution and 250 MS/s sampling rate. The dynamic
range is configured to be 2.0 Vpp, leading to 0.122 mV for the least
significant bit (one ADC count). In PandaX-4T, the typical single
photoelectron (SPE) signals from 3-inch PMTs have the amplitude around 7
mV and the pulse width around 20-30 ns. An example of the digitized
waveform from a SPE signal is shown in Figure.~\ref{spe_wf} left.

\begin{figure*}[!htbp]
  \centering
  \includegraphics[width=0.8\linewidth]{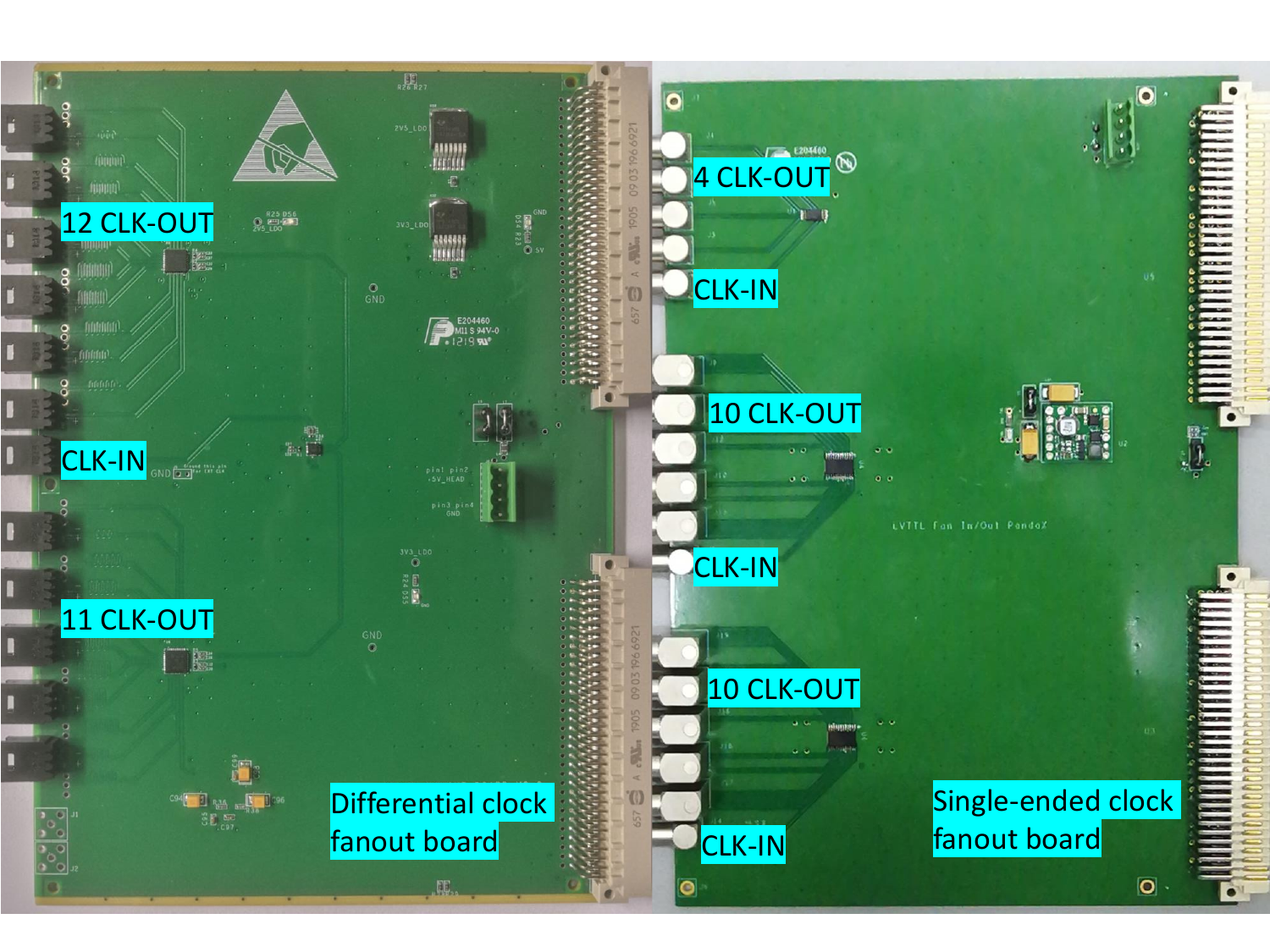}
  \caption{The differential (left) and single-ended (right) clock
    fanout module used for the PandaX-4T digitizer system. The former
    provides synchronous external clocks for the digitizers. The latter provides synchronous
    start-acquisition signals and external-trigger signals.
  }
  \label{fanout}
\end{figure*}

As mentioned in the introduction, a triggerless readout scheme is used
in the PandaX-4T experiment.  This is achieved by using the dynamic
acquisition window (DAW) algorithm implemented in the V1725
digitizers~\cite{Ref:dawmanual}. On a channel-by-channel basis,
digitized samples above a threshold can be automatically identified.
This is referred as the self-trigger. Then a pre-configured number of
data (before and after the trigger time) are saved into the buffer in
the digitizer for readout, together with other information such as the
trigger time tag and the channel number. Figure.~\ref{spe_wf} right
shows a typical recorded amplitude distribution from a 3-inch PMT,
with the self-trigger threshold set to be 20 ADC counts.  The typical
amplitude of a SPE signal is about 60 ADC counts, so this threshold
corresponds to approximately 1/3 PE.  To ensure timing synchronization
among all digitizers, common external clock signals are used.  These
clock signals are originated from a common clock oscillator and distributed by
custom designed differential clock fanout modules (see Figure~\ref{fanout} left). In this way, there is
no need to tune the clock phase as in the previous electronics
system~\cite{Ren:2016ium}, where clock signals were daisy-chained
among all digitizers. The start-acquisition of the digitizer is configured to be ``first trigger controlled''
~\cite{Ref:dawmanual}. The acquisition is started once receiving a pulse originated from a generic logic unit
(CAEN V1495) and distributed by custom designed single-ended clock
fanout modules (Figure~\ref{fanout} right).

In addition to the self-trigger mode, the digitizers can be configured
to acquire data with external triggers. In this mode, a number of data
with equal length from all channels are read out without baseline
suppressions. This can be used for PMT gain calibrations and
self-trigger efficiency studies.  

The last part includes the data acquisition, processing, and
storage. Four readout DAQ servers (Dell R730) are used to acquire data
from the 32 digitizers. Each DAQ server is connected with 8 digitizers
through individual optical fibers with two PCIE interface cards (CAEN
A3818C). The data transfer rate is up to 85 MB/s for each
digitizer. The acquired data of each server are sent to another data
server (Dell R930) via a 10 Gbps optical fiber switch. In the data
server, all received data are sorted according to the trigger time tag
and written into disk afterwards.  Raw data are transferred to a
computer cluster for offline processing and analysis.

In PandaX-4T, the DAQ system is controlled via a web-based
interface. Before each run, parameters such as the trigger mode (self
or external trigger), self-trigger threshold of each channel, record
length of each trigger, can be configured on this interface and are
written into a database.  After the DAQ system is started, the readout
DAQ servers will configure the digitizers using the parameters from
the database. The above-mentioned start-acquisition pulse is sent to
all digitizers only after they have been configured, so all digitizers
start to acquire data at the same time.

\section{Performance}

\begin{figure}[!htbp]
  \centering \includegraphics[width=0.48\textwidth]{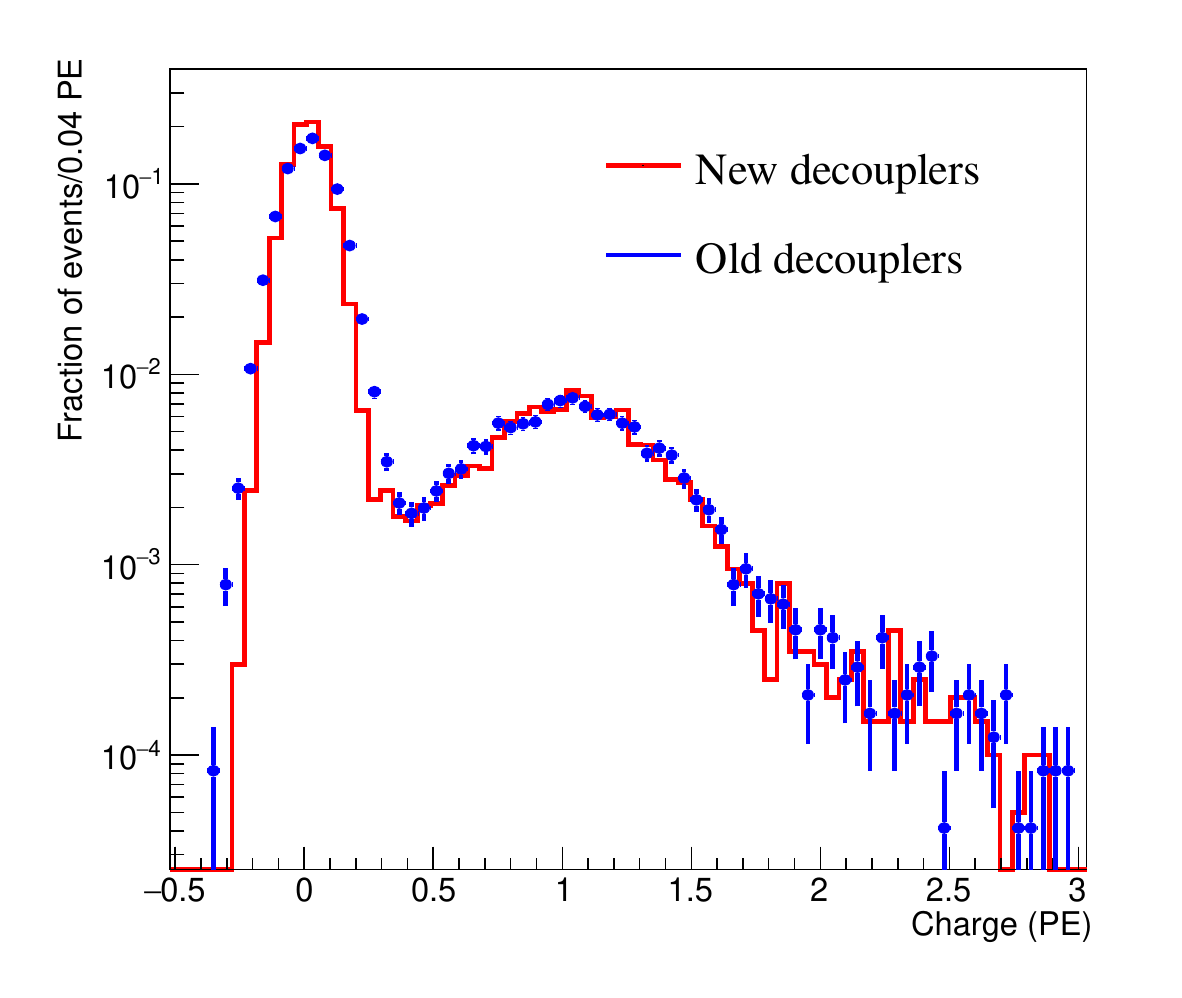}
  \includegraphics[width=0.48\textwidth]{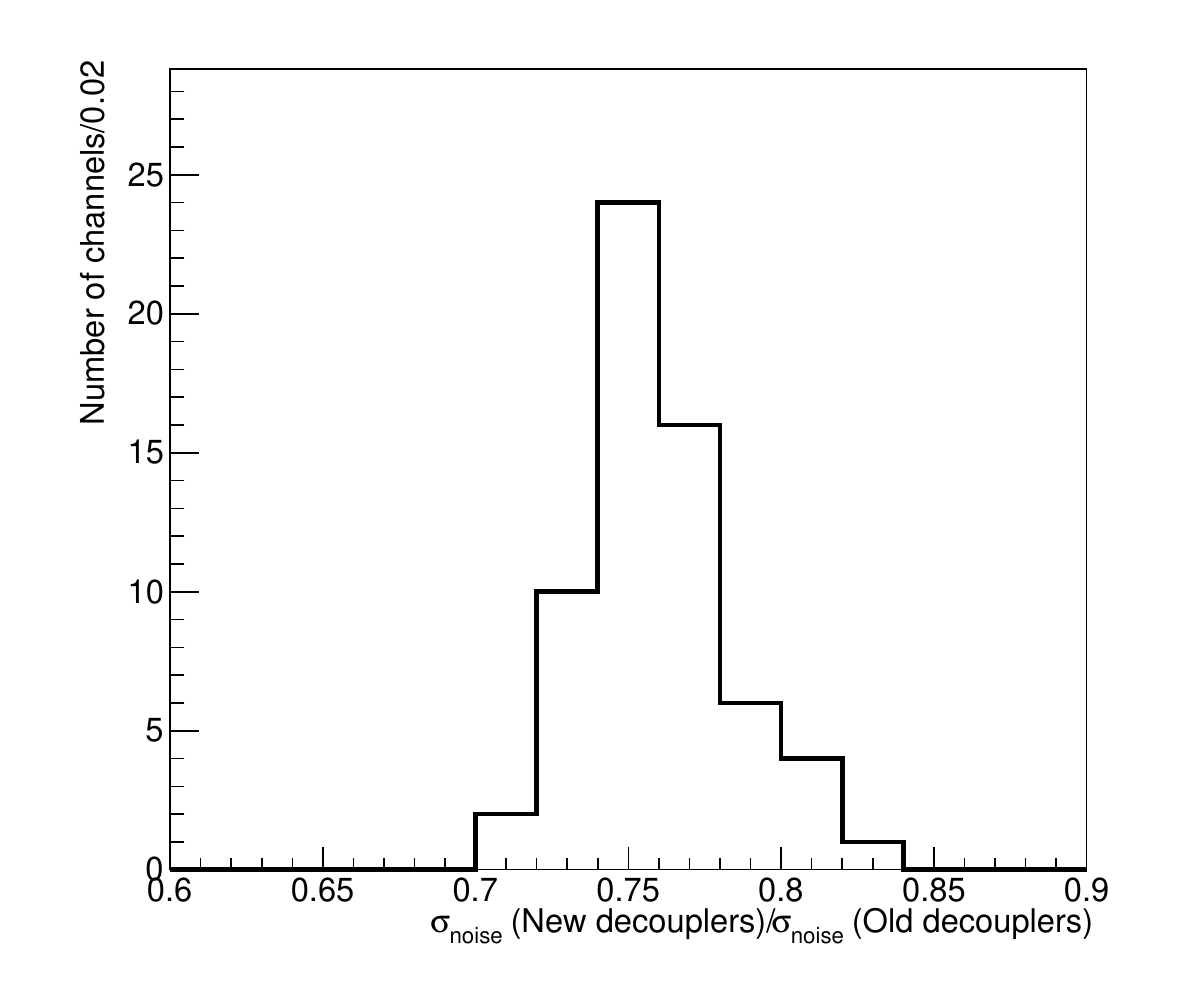}
  \caption{Left, comparison of the charge distributions of one 3-inch
    PMT, with the new (red) and the old (blue) decoupler
    module. Right, the ratio of the pedestal width, measured using 64
    new decouplers and 64 old decouplers.}
  \label{newdc}
\end{figure}

We start by showing the performance of the new decoupler
modules. Using low-intensity LED calibration data, we measured the SPE
charge distributions for 64 PMTs with 64 old decouplers. The same measurements are
repeated with 64 new decouplers at the same
condition. Figure~\ref{newdc} left shows the comparison of the charge
distributions from one PMT. The peak around zero is the pedestal
region and corresponds to electronic noise in the baseline.  The
second peak corresponds to the PMT response to a single photoelectron. As
expected in last section, the chosen amplifiers have little impact on
the charge measurement of  signals.  But there is a reduction in the pedestal width of
24\% on average with the new decouplers, shown in Figure~\ref{newdc} right. This corresponds to a
 32\% improvement on the signal-over-noise ratio. 

\begin{figure}[!htbp]
  \centering
  \includegraphics[width=0.48\textwidth]{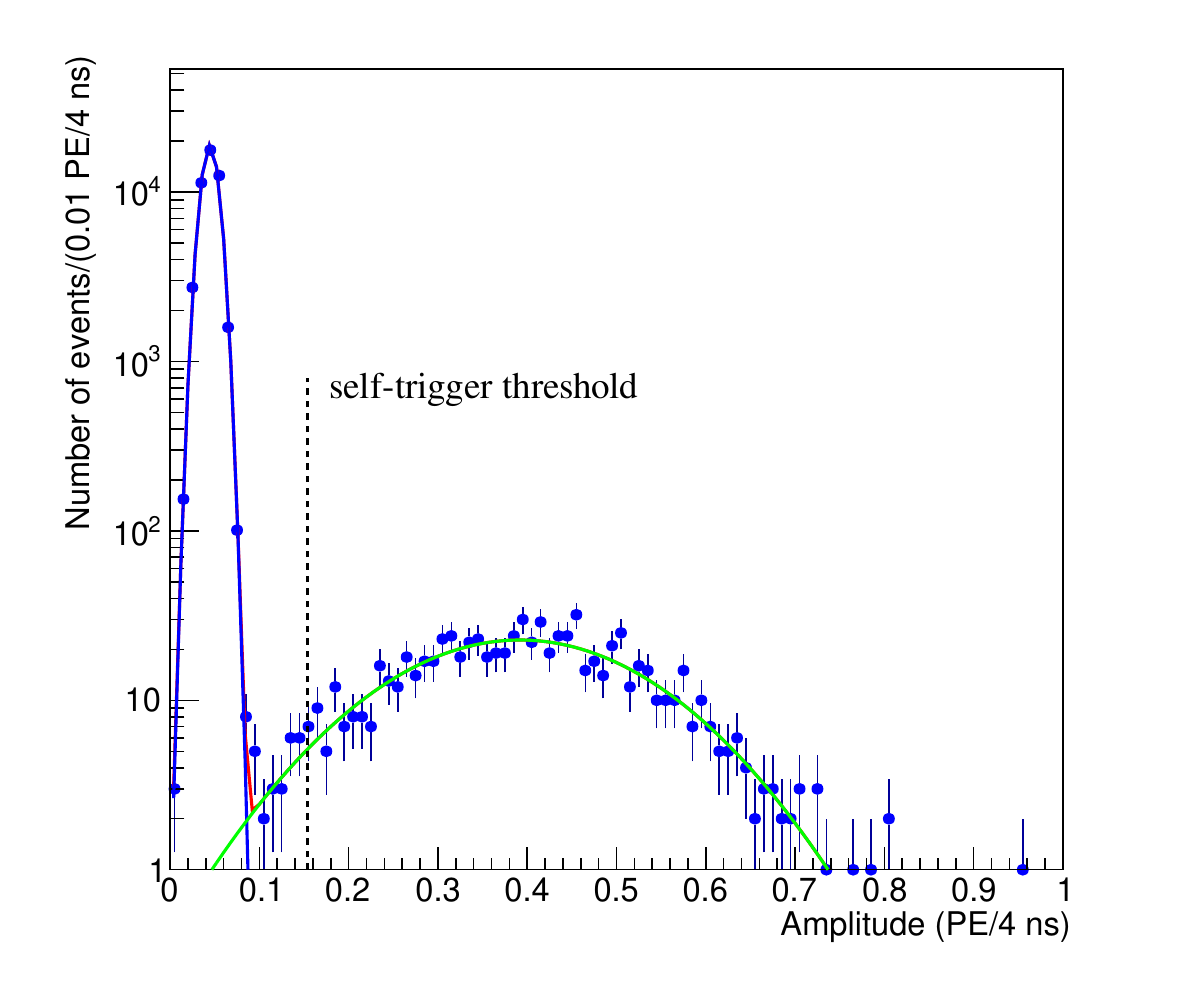}
  \includegraphics[width=0.48\textwidth]{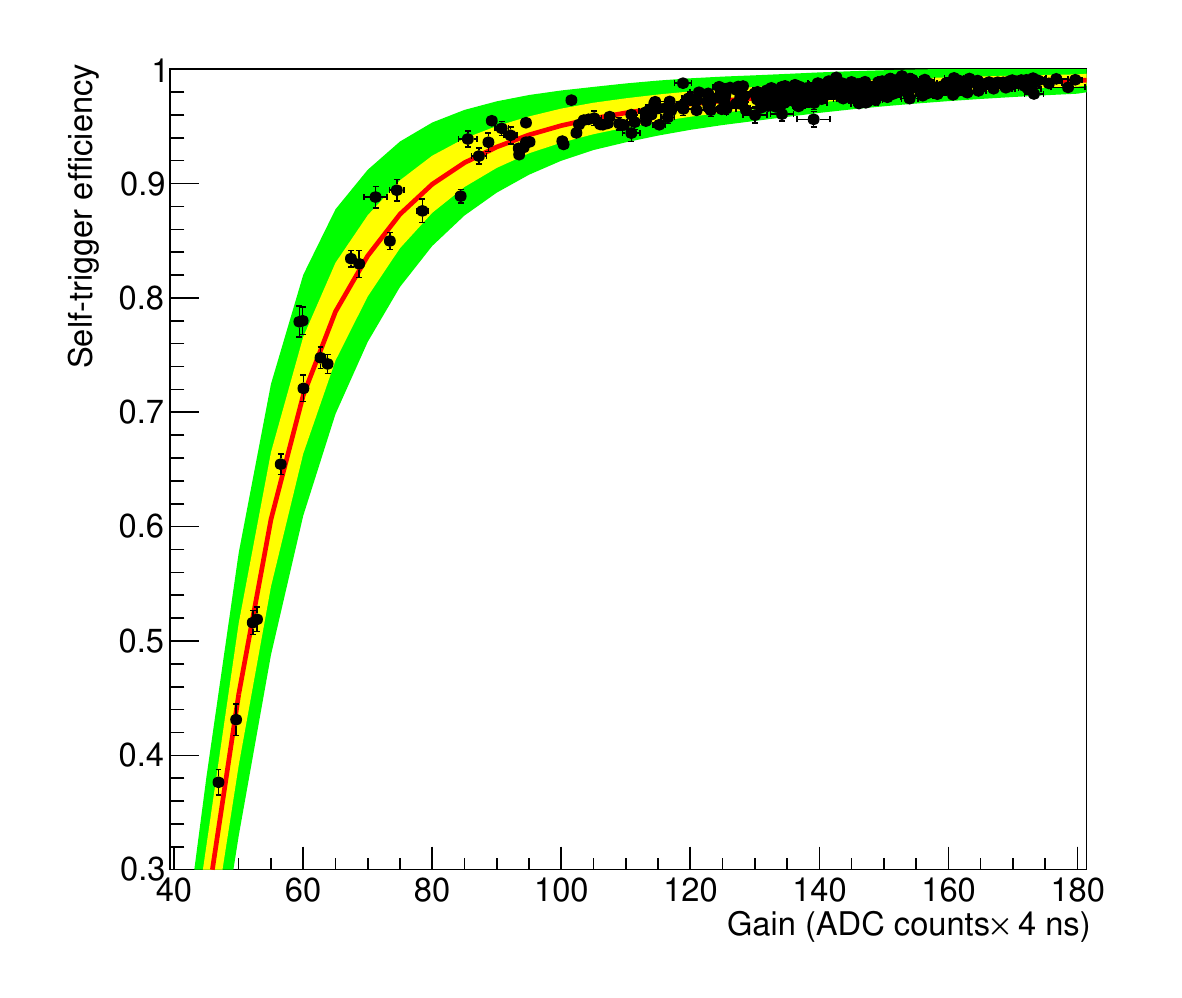}
  \caption{Left, amplitude spectrum (in unit of PE/4 ns) of a R11410 3-inch
    PMT in low-intensity LED calibration data. Superimposed function
    (red) is a combined fit of two Gaussian functions to model the
    noise (blue) and the SPE signals (green). The black-dashed line
    denotes the self-trigger threshold for this channel. Right,
    self-trigger efficiency of SPE signals vs. gain for each 3-inch
    PMT channel. The data points are measurements. The red curve and
    the bands are predictions and uncertanties using inputs from
    measurements. See text for details.  Here, the gain refers to the
    mean of the integral of raw digitized waveform for SPE pulses.
    A gain of 100 ADC counts$\times$ 4 ns corresponds to a PMT gain of
    approximately $4\times10^{6}$.}
  \label{eff}
\end{figure}

Then we show the efficiencies of the self-trigger threshold on SPE
signals.  This is one of the key parameters that affects the signal
efficiency in the triggerless readout scheme. During the commissioning
runs of PandaX-4T, the threshold is set to be 20 ADC counts for 3-inch
PMTs. The efficiency of recording SPE signals is estimated using low
intensity LED calibration data without baseline suppression applied in
the digitizers.  These data are recorded with 50 Hz external triggers
that are synchronous to the pulse which drives the LED. So a possible SPE signal can be searched for in
a fixed window of each recorded data. The resultant amplitude (in unit
of PE/4 ns) of one 3-inch PMT is shown in Figure~\ref{eff} left. In
this plot, the raw amplitude (in unit of ADC counts) is divided by the
gain of this PMT channel determined using light calibration data.  Here, the
gain is in unit of ADC counts$\times$4 ns (as in Figure 7 right), where
4 ns corresponds to the sampling interval of the digitizers. The
distribution is fitted with a sum of two Gaussians to describle the
noises and signals, respectively. The SPE amplitude mean is about 0.39
PE/4 ns, with a relative variation of 4\% among all channels. The mean
values have almost no dependence on the gain. On the other hand, the SPE amplitude width varys from 0.09 PE/4 ns to 0.12 PE/4 ns as the gain increases, with a relative variation of 8-11\% among
channels. 

The SPE trigger efficiency is estimated as the fraction of signal
events above the threshold. Figure.~\ref{eff} right shows the obtained
efficiency vs. the gain of each 3-inch PMT channel. Using the
above-mentioned mean value of SPE amplitude of all channels, and the
gain-dependent amplitude width, we can predict the efficiency for any
given gain. This is the red curve in Figure.~\ref{eff} right.  Taking
into account the above-mentioned variations among channels, we can
estimate how much the efficiencies are expected to change.  This is
shown as the yellow band. If we consider twice of the variations, the
efficiencies vary within the green band. Overall, we can see that the
measurements are consistent with the prediction. The average
self-trigger efficiency of all 3-inch PMTs is 96\%. The above
measurements have been used to model the S1-loss in the signal model
for the first DM search~\cite{p4paper}.

\begin{figure*}[!htbp]
  \centering
  \includegraphics[width=0.48\textwidth]{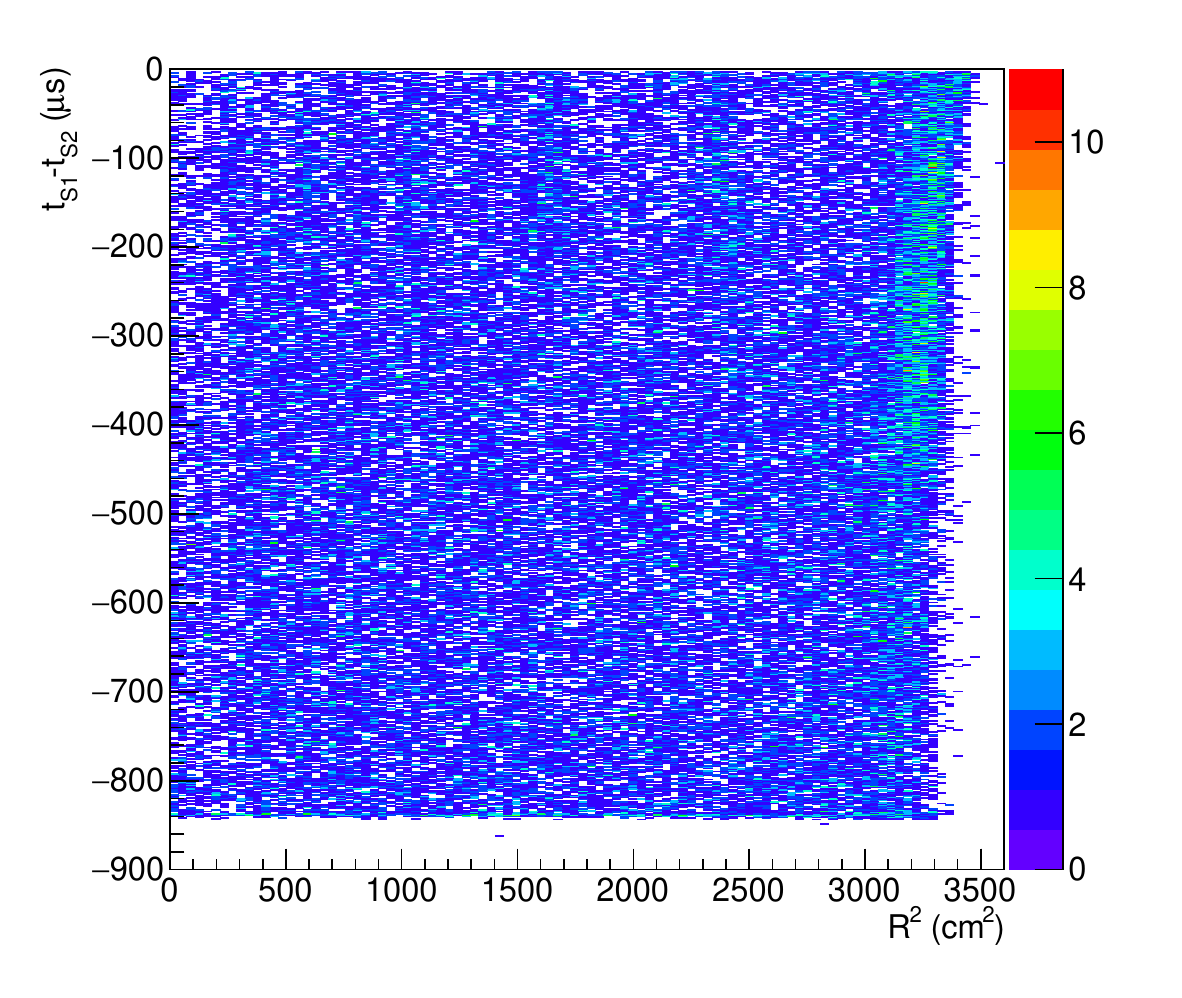}
  \includegraphics[width=0.48\textwidth]{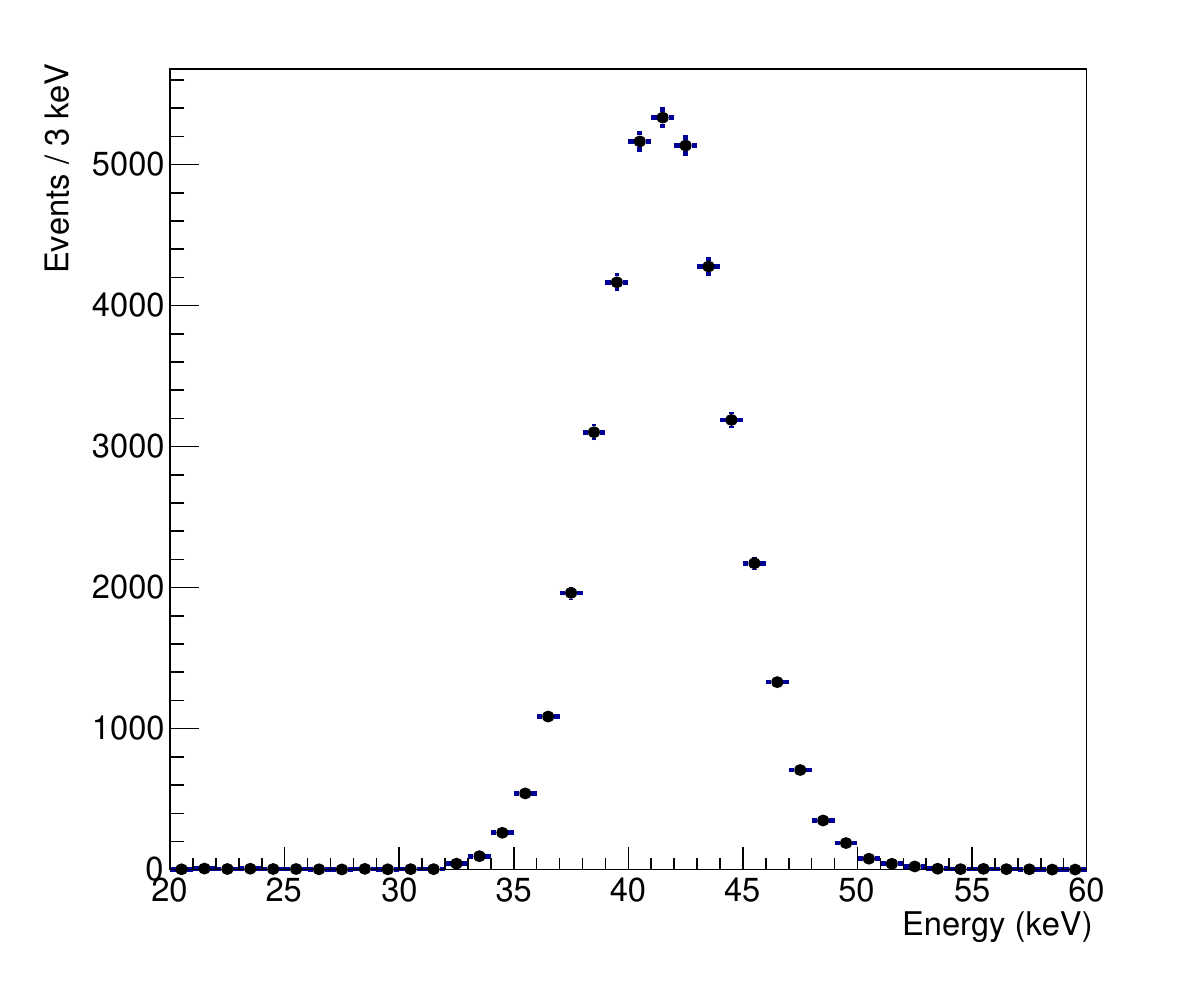}
  \caption{Left, vertex distribution of events with reconstructed
    energy between 20 and 60 keV in $^{83m}$Kr calibration
    data. Right, the reconstructed energy spectrum of the two
    conversion electrons with total energy 41.5 keV from $^{83m}$Kr
    decays.}
  \label{spectrum}
\end{figure*}

\begin{figure*}[!htbp]
  \centering \includegraphics[width=0.95\textwidth]{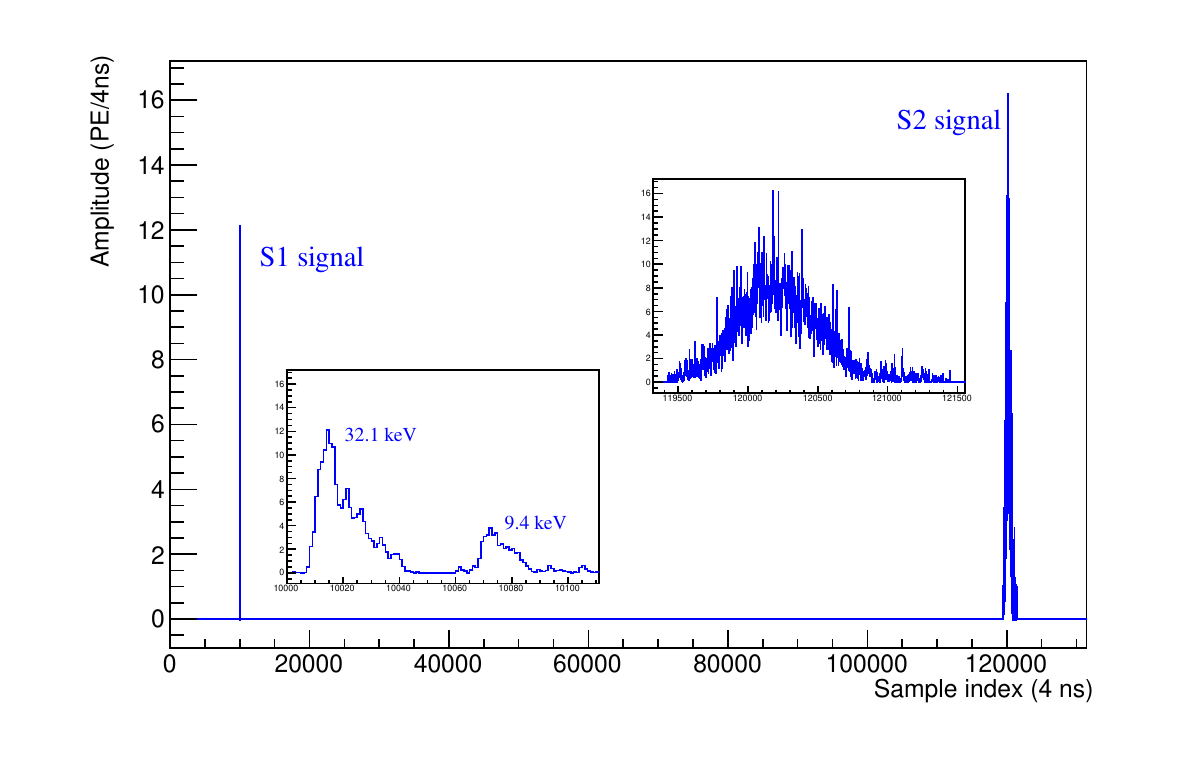}
  \caption{A typical waveform from $^{83m}$Kr decays. The S1 signal
    contains two pulses. The first one corresponds to the emission of
    the conversion electron with energy of 32.1 keV. The second one
    corresponds to the 9.4 keV conversion electron, emitted following
    the first electron with a 154 ns half-life.}
  \label{event_wf}
\end{figure*}

Unlike previous PandaX experiments, there is no more definition of
events at the DAQ level in PandaX-4T. At the DAQ level, waveform
information is recorded at the channel-by-channel basis and
independently among all channels. Only in offline analysis, S1 and S2
signals are identified using information from all
channels. Afterwards, a physical event is built by combining S1 and S2 signals
within a window of 1 ms. We use physics data to validate the DAQ.
Unless otherwise stated, all data used here and
below were recorded during the commissioning phase of the detector
from Nov 2020 to May 2021. The cathode of the TPC was applied with a
HV of -16 or -18 kV.  The drift velocity of electrons in the liquid is about
2 mm$/\mu$s, so a maximum electron drift time of about 800 $\mu$s is
expected. The gate of the TPC is applied with a HV of -5 kV to
extract the electrons into the gaseous region. The measured electron
extraction efficiency is above 90\%~\cite{p4paper}. Then the deposited
energy in the liquid xenon from this event can be determined from the
S1 and the S2 signals, together with other detector parameters
including photon detection efficiency, electron extraction efficiency,
the single electron gain, and other corrections~\cite{p4paper}.
Figure~\ref{spectrum} left shows the vertex distribution of selected
events with reconstructed energies between 20 and 60 keV from
$^{83m}$Kr calibration data. A maximum drift time of 840 $\mu$s is
observed, consistent with expectation. Figure~\ref{spectrum} right
shows the reconstructed energy spectrum which corresponds to the two
conversion electrons with expected total energy of 41.5 keV from
$^{83m}$Kr decays. Events with only one pair of S1 and S2 signals are
selected.  The event vertex is required to satisfy $R^{2}<2500$
cm$^{2}$ and 20 $\mu$s $< t_{S2}-t_{S1}<700~\mu$s to reduce external
backgrounds. The S1 and S2 waveforms of one selected event is shown in
Figure~\ref{event_wf}. In this event, the two S1-like pulses from the
cascade decays of $^{83m}$Kr are not well separated and thus merged as
one S1 signal. These results validate the electronics and DAQ system.

In PandaX-4T, the DAQ system is designed to save all self-triggered
data in the digitizers to disk for offline analysis. This imposes
larger bandwidth requirement compared to the global-trigger based DAQ
system. A maximum bandwidth of 470 MB/s was achieved with a Pu-C
neutron source placed outside the TPC during an earlier test run in
2020. During the commissioning runs, the bandwidths of most runs range
from 20 MB/s to 80 MB/s, depending on the run conditions. A bandwidth
of less than 100 MB/s is also preferred for two reasons. The effect of
data loss due to digitizer busy is negligible as discussed
below. During the data acquisition, the raw data are transferred from
CJPL-II to the computer cluster in Chengdu through a 1 Gbps optical
link. The maximum bandwidth is about 100 MB/s.
  
\begin{figure}[!htbp]
  \centering \includegraphics[width=0.95\textwidth]{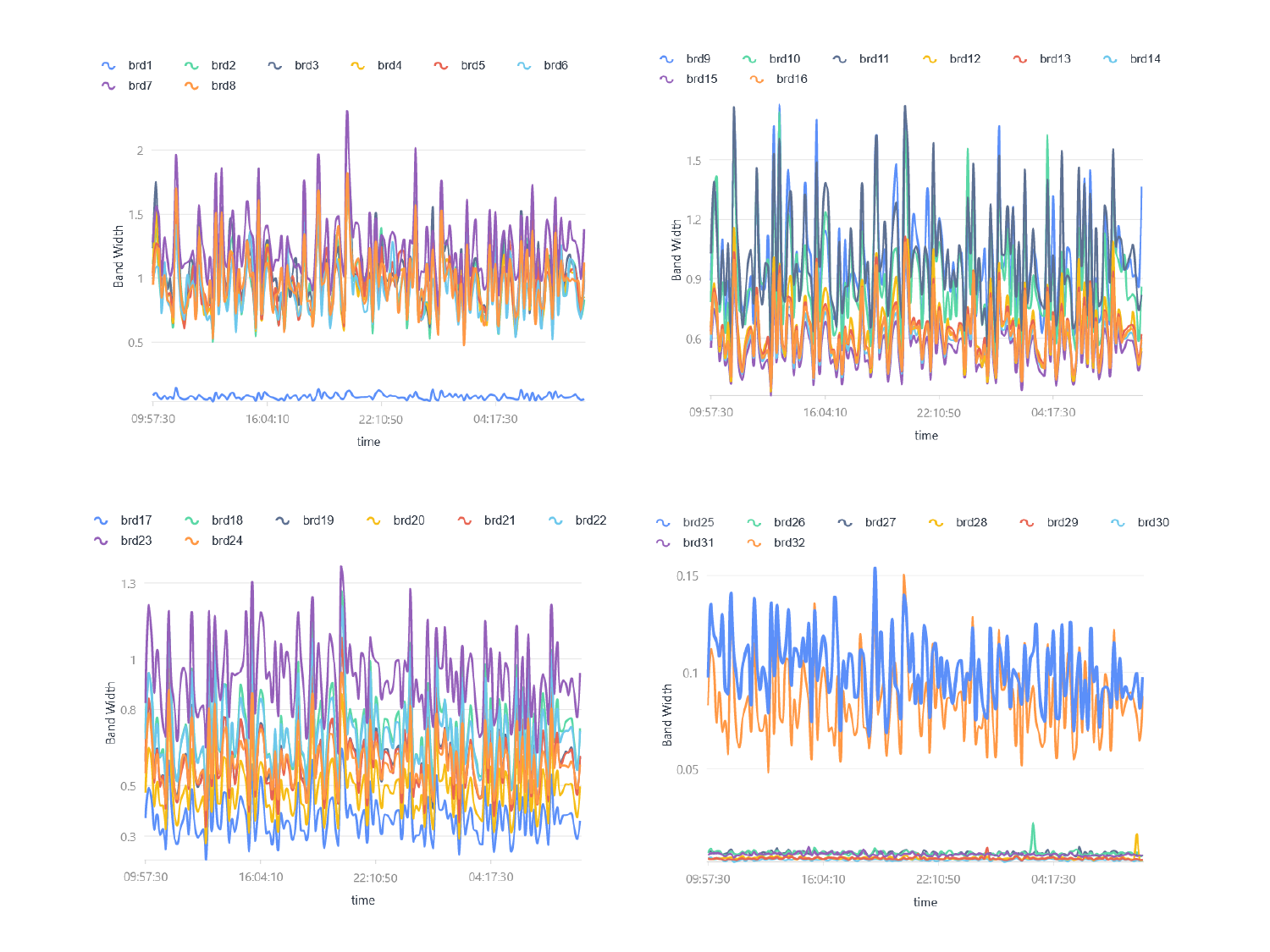}
  \caption{The bandwidth of each digitizer (unit: MB/s) as a function of time
    during a typical background run for DM searches during the commissioning runs of PandaX-4T.}
  \label{bandwidth}
\end{figure}

\begin{figure*}[!htbp]
  \centering \includegraphics[width=0.48\textwidth]{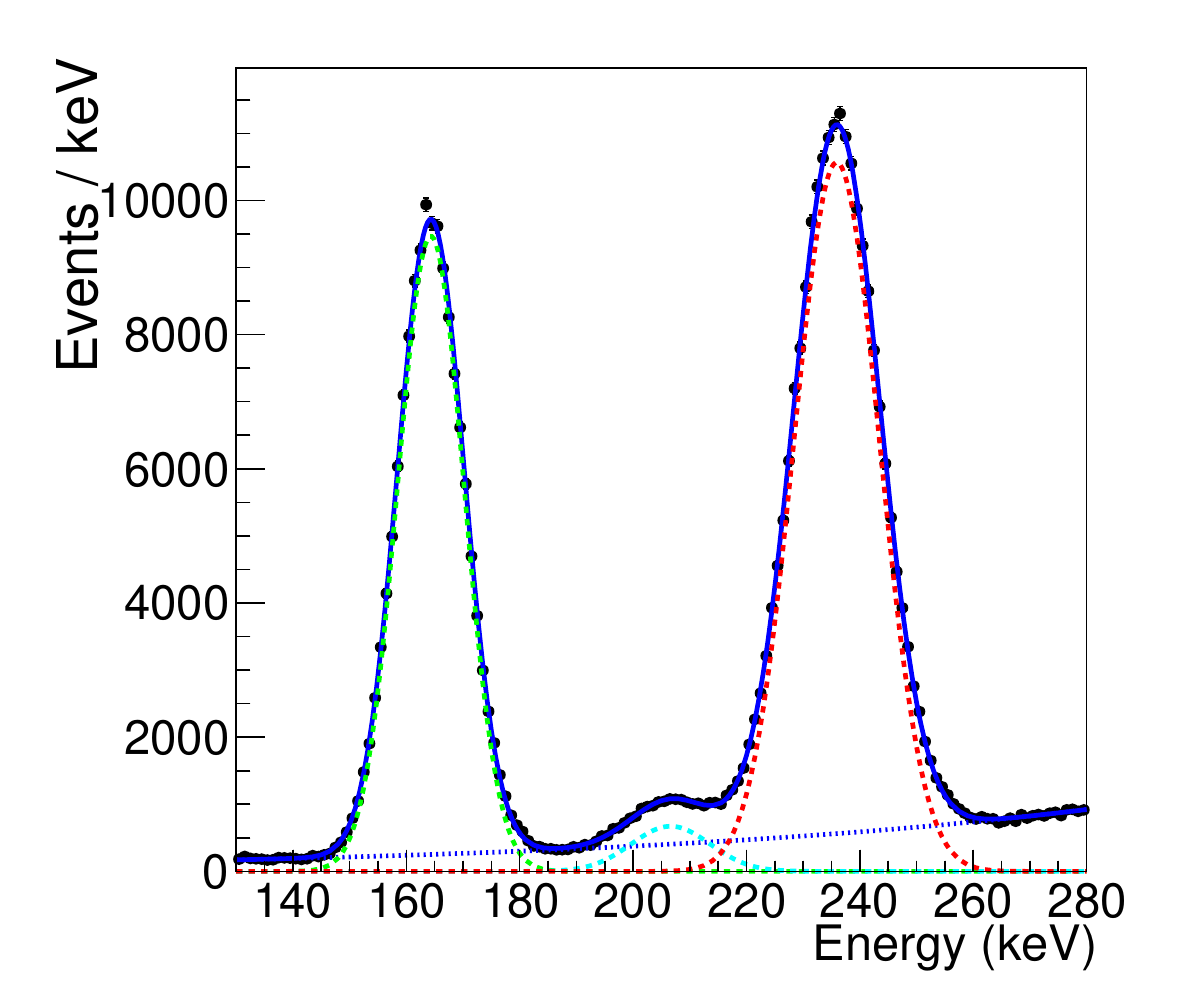}
  \caption{Reconstructed energy spectrum from 130 keV to 280 keV in
    background runs. The distribution is fitted with a sum of three
    Gaussian functions to model the $\gamma$ peaks, and an exponential
    function to model the continous background.}
  \label{fit164}
\end{figure*}

Finally, we discuss the effect of data loss due to the digitizer
busy. This might cause nonnegligible systematic errors in the event reconstruction. 
During the background runs for DM searches, the
bandwidth is only about 20 MB/s, with each digitizer contributing up to
1.5 MB/s, shown in Figure~\ref{bandwidth}.  The busy time of the
digitizers at this level of the bandwidth is negligible, given the
on-board buffer of 10 MB/channel and the readout bandwidth limit of 85
MB/s per digitizer. This is verified by comparing the reconstructed
peaks of known $\gamma$ rays in background runs and calibration
runs. Figure~\ref{fit164} shows the several high energy reconstructed $\gamma$
peaks in background runs. Here and below, the above-mentioned event and vertex selections are applied. 
The two peaks with largest statistics correspond to the 164
keV and 236 keV $\gamma$ rays from $^{131m}$Xe and $^{129m}$Xe decays,
respectively. The two peak values in data are determined to be 164.32$\pm$0.02 keV and
235.92$\pm$0.02 keV. A number of runs were taken with neutrons
injected into the TPC from a deuterium-deuterium (D-D) fusion
source. The bandwidths range from 60 to 80 MB/s, with contribution from
each digitizer up to 5-6 MB/s. The two peak values in these data are
163.8$\pm$0.1 keV and 235.1$\pm$0.1 keV. The difference compared to
background runs is 0.3\%. This is consistent with the 0.3\% variation
of the 164 keV peak values obtained in different background runs~\cite{MGtalk}. This
shows that the impact of digitizer busy is negligible in these data
acquisition conditions.

One run was taken with a $^{137}$Cs source placed at the DD injection
tunnel.  The bandwidth is 120 MB/s. The maximum bandwidth from single
digitizer is 11 MB/s. The two peaks become 161.7$\pm$0.2 keV and
233.6$\pm$0.4 keV. The maximum difference compared to background runs
is 1.6\%. This illustrates that digitizer busy causes noticeable
systematic effect in this run condition.  Therefore, it might be
useful to use the global triggers to acquire data in this case.
However, due to the limitation of the digitizers, S1 and S2 signals
can not be recorded in one trigger. Only data less than several $\mu$s
in advance of the trigger time can be recorded. This problem can be
solved with a new type of custom designed digitizers~\cite{adc}.

\section{Summary}

In summary, we presented the electronics and the DAQ system in the
PandaX-4T experiment. Waveform of each PMT is
recorded if the pulse is above the threshold on a channel-by-channel
basis. The average efficiency of recording SPE signals of the 3-inch
PMTs is 96\%. The DAQ system is designed to save all recorded data in
the digitizers for offline analysis. The maximum bandwidth of the DAQ
system is above 450 MB/s. This represents an improvement of more
than a factor of 6 compared to the system in previous PandaX
experiments~\cite{Ren:2016ium}. Many DM searches are expected
to benefit from this triggerless DAQ, since there is no more
inefficiencies due to the global triggers. In the first DM search in
PandaX-4T~\cite{p4paper}, the lower threshold on S2 signal
is already reduced to 80 PE from 100 PE which was used in the PandaX-II
final WIMP analysis~\cite{{Wang:2020coa}}. The presented system has been used to
successfully acquire data during the PandaX-4T commissioning phase
from Nov 2020 to May 2021. Data taking will be resumed after a
tritium removal campagain.

\section{Acknowledgement}
This project is supported by grants from the Ministry of Science and Technology
of China (No. 2016YFA0400301 and 2016YFA0400302), a Double Top-class grant from
Shanghai Jiao Tong University, grants from National Science Foundation of China
(Nos. 11875190, 11505112, 11775142 and 11755001), supports from the Office of
Science and Technology, Shanghai Municipal Government (18JC1410200), and support
also from the Key Laboratory for Particle Physics, Astrophysics and Cosmology,
Ministry of Education. This work is supported also by the Chinese Academy of
Sciences Center for Excellence in Particle Physics (CCEPP).


\begin{thebibliography}{1}

\bibitem{Ren:2016ium}
  X.~Ren et~al.,
  \newblock The Electronics and Data Acquisition System for the PandaX-I Dark Matter Experiment,
  \newblock 2016 {\em JINST} {\bf 11} T04002

\bibitem{Wu:2017cjl}
  Q.~Wu et~al.,
  \newblock Update of the trigger system of the PandaX-II experiment,
  \newblock 2017 {\em JINST} {\bf 12} T08004

\bibitem{Ren:2018gyx}
  PandaX-II Collaboration,
  \newblock Constraining Dark Matter Models with a Light Mediator at the PandaX-II Experiment,
  \newblock {\em Phys. Rev. Lett.} {\bf 121} (2018) 021304

\bibitem{Yang:2021adi}
  PandaX-II Collaboration,
  \newblock Constraining self-interacting dark matter with the full dataset of PandaX-II,
  \newblock {\em Sci. China Phys. Mech. Astron.} {\bf 64} (2021) 111062.
  
\bibitem{Cheng:2021fqb}
  PandaX-II Collaboration,
  \newblock Search for Light Dark Matter-Electron Scatterings in the PandaX-II Experiment,
  \newblock {\em Phys. Rev. Lett.} {\bf 126} (2021) 211803

\bibitem{Zheng:2020kfp}
  Q. Zheng, et~al.,
  \newblock An improved design of the readout base board of the photomultiplier tube for future PandaX dark matter experiments,
  \newblock 2020 {\em JINST} {\bf 15} T12006

\bibitem{adi} AD8009
  \newblock 1 GHz, 5500 V/$\mu$s Low Distortion Amplifier,
  \newblock \url{https://www.analog.com/en/products/ad8009.html}
  
\bibitem{Ref:dawmanual} DPP-DAW,
\newblock Digital Pulse Processing with Dynamic Acquisition Window,
\newblock \url{https://www.caen.it/products/dpp-daw/}

\bibitem{p4paper} PandaX-4T Collaboration, 
  \newblock Dark Matter Search Results from the PandaX-4T Commissioning Run,
  \newblock {\em Phys. Rev. Lett.} {\bf 127} (2021) 261802

\bibitem{MGtalk} J. Liu, 
\newblock First Results from PandaX-4T, talk given in the 16$^{th}$ Marcel Grossmann meeting,
\newblock \url{https://indico.icranet.org/event/1/}.

\bibitem{adc} C. He et~al.,
\newblock  A 500 MS/s waveform digitizer for PandaX dark matter experiments,
\newblock 2021 {\em JINST} {\bf 16} T12015

\bibitem{Wang:2020coa}
  PandaX-II Collaboration,
  \newblock Results of dark matter search using the full PandaX-II exposure,
  \newblock {Chin. Phys. C} {\bf 44} (2020) 125001
  
\end{thebibliography}
\end{document}